\documentclass[reprint,aps,prd,floats,floatfix,amsmath,amssymb,amsfonts,nofootinbib,longbibliography,superscriptaddress]{revtex4-1}

\usepackage{graphicx}
\usepackage{dcolumn}
\usepackage{bm}
\usepackage{tensor} 
\usepackage{natbib}
\usepackage{subcaption}

\usepackage[hyperindex,colorlinks]{hyperref}

\newcommand{\s}{\scriptscriptstyle}

\begin{document}

\title{Newtonian-like gravity with variable $G$}

\author{J\'ulio C. Fabris}
\email{julio.fabris@cosmo-ufes.org}%
\affiliation{%
N\'ucleo Cosmo-ufes \& Departamento de F\'isica,  Universidade Federal do Esp\'irito Santo (UFES)\\
Av. Fernando Ferrari, 540, CEP 29.075-910, Vit\'oria, ES, Brazil.}%
\affiliation{%
National Research Nuclear University MEPhI, Kashirskoe sh. 31, Moscow 115409, Russia}%
\author{Tales Gomes}%
\email{talesaogomes@hotmail.com}
\affiliation{%
N\'ucleo Cosmo-ufes \& Departamento de F\'isica,  Universidade Federal do Esp\'irito Santo (UFES)\\
Av. Fernando Ferrari, 540, CEP 29.075-910, Vit\'oria, ES, Brazil.}%

\author{J\'unior D. Toniato}%
\email{junior.toniato@ufop.edu.br}
\author{Hermano Velten}%
\email{hermano.velten@ufop.edu.br}
\affiliation{%
Departamento de F\'isica, Universidade Federal de Ouro Preto (UFOP), Campus Universit\'ario Morro do Cruzeiro, 35.400-000, Ouro Preto, Brazil}%
\date{\today}

\begin{abstract}
We propose a Lagrangian formulation for a varying $G$ Newtonian-like theory inspired by the Brans-Dicke gravity. Rather than imposing an {\it ad hoc} dependence for the gravitational coupling, as previously done in the literature, in our proposal the running of $G$ emerges naturally from the internal dynamical structure of the theory. We explore the features of the resulting gravitational field for static and spherically symmetric mass distributions as well as within the cosmological framework.
\end{abstract}

\maketitle

\section{Introduction}

Physics is based on a set of fundamental constants and each one of it can be related with specific phenomena. For instance, gravity is associated with the constant $G$, quantum mechanical effects are related with the Planck's constant $\hbar$, the speed of light $c$ refers to relativistic effects and thermodynamic processes are linked to the Boltzmann constant $k_B$. When two or more of such constants appear in an equation one can associate it to a characteristic area of physics e.g., relativistic quantum mechanics makes use of $\hbar$ and $c$. By including $G$ to the latter analysis one deals with typical quantum gravity effects. The study of the black hole thermodynamics as a quantum effect appearing in gravitational systems should contain all constants mentioned above.

Although $G$ has been the earliest constant introduced in a physical theory, its value is the least accurate in comparison with the other constants. Up to date $G$ measurements still have uncertainties of order of $10^{-4}$ \cite{Will:2014kxa}. This is deeply related to the fact that all experimental apparatus designed to measure the gravitational constant can not completely eliminate the gravitational interaction from exterior masses due to the universal attractive long range behavior of gravity.

There are different approaches to measure $G$ as for example using microgravity environment (see proposals in \cite{PhysRevLett.116.231101}), free-fall methods \cite{Schwarz2230}, beam balance \cite{PhysRevD.74.082001}, simple pendulum experiments \cite{PhysRevLett.105.110801} or atom interferometry \cite{2008mgm..conf.2519B,Rosi_2016}. Independently of the specific technique used all such experiments are performed in an almost zero curvature spacetime in which the Newtonian limit of any covariant gravitational theory does apply. 

In laboratory experiments any deviation of the gravitational inverse square law or even a long-time cosmological dependence of the gravitational coupling may be detected by fitting some kind of parameterization of $G$. How such parameterization should look like? For example, in the search of a new gravitational interaction feature as a Yukawa like correction $G(r)=G_{\infty} (1+\mu e^{-r/\lambda})$, where the parameters $\mu$ (the magnitude of this correction) and $\lambda$ (characterizing the typical range of the Yukawa correction) have to be fitted by the experiment outcomes. The constant $G_{\infty}$ is the long range gravitational coupling i.e., $G(r \gg \lambda)\approx G_{\infty}$. The Yukawa correction is well motivated since it is present in massive scalar-tensor theories of gravity, a commonly considered alternative to general relativity. However, different corrections to the standard inverse square law may be allowed. But what is the physical origin of such possible deviations? In the context of modified gravity theories the weak field limit shows how such a theory behaves with respect to local gravitational tests. In case new predictions are not satisfied by e.g., the solar system tests, one still can recover the classical general relativistic results by evoking screening mechanisms. The latter suppress any new gravitational feature in specific environments such as the Earth or the solar system. Actually, such mechanisms act assuring that nonlinearities can be important on cosmological scales, guaranteeing an accelerated background expansion. But they are irrelevant for local systems. Once that, different gravitational theories may involve different screening mechanisms \cite{Koyama:2015oma,Brax_2013,Khoury:2013yya,Babichev_2013,Koyama:2015vza,Ishak:2018his}, in our point of view, a Newtonian theory with varying $G$ may shed light on some features of specific screening mechanisms by introducing variable-$G$ effects starting with a formal Newtonian gravitational Lagrangian. Even if this issue is not the central aspect of our analysis, we can obtain some hints on how to treat this problem with a varying $G$ theory. 


Since the possible time variation of $G$ is well constrained by Lunar Laser Ranging experiments as being of order $\dot{G}/G \lesssim 10^{-13} yr^{-1}$ \cite{PhysRevLett.93.261101}, laboratory experiments are not designed to probe the time variation of $G$ on human time scales and, therefore, this is expected to occurs on cosmological time-scales. However, there many proposals for such dependence.
Even though the discussion of a time varying $G$ has been introduced by Milne in 1935 \cite{1935rgws.book.....M}, the possibility of a variable gravitational coupling is usually attributed to a Dirac's discussion, motivated by a supposed coincidence appearing when combining some physical constants \cite{Dirac:1937ti, Dirac:1938mt}. Considering the ratio between the electric and gravitational forces at atomic scale, one has 
\begin{eqnarray}
\frac{F_g}{F_e} = 4\pi \epsilon_0 \frac{Gm_pm_e}{e^2} \sim 10^{-40},
\end{eqnarray}
where $m_p$ and $m_e$ are the masses of proton and electron, respectively, $e$ is the fundamental electric charge and $\epsilon_0$ the vacuum permittivity. Defining an atomic time scale $t_A = e^2/4\pi\epsilon_0 m_e c^3 \sim 10^{-23}$\,s, and comparing it with the universe's age, associated with the Hubble constant such that $t_H = H_0^{-1} \sim 1,7\times 10^{17}$\,s, it is once again obtained,
\begin{eqnarray}
\frac{t_A}{t_H}  \sim 10^{-40}.
\end{eqnarray}

Another coincidence pointed out by Dirac is that a specific combination of $G$, $c$, $\hbar$ and $H_0$ gives the pion mass, namely
\begin{eqnarray}
\label{coin}
\biggr(\frac{\hbar^2 H_0}{Gc}\biggl)^\frac{1}{3} \sim m_\pi.
\end{eqnarray}
A detailed discussion of such coincidences is found in Ref.~\cite{Weinberg:1972kfs}. Relation (\ref{coin}) is the source of the original argument used by Dirac. If relation (\ref{coin}) really expresses some fundamental aspect of nature, and since $H_0$ is related to the present age of the Universe, other fundamental constant must vary with time. The natural choice is to assume that $H/G$ remains constant along universe lifetime. Given that $H \sim t^{-1}$ the same time behaviour should apply to $G$, i.e. $G \propto t^{-1}$.

Albeit Dirac's argument is not as solid as one would expect, it motivated researchers to investigate the nature of physical constants and whether they are truly invariant quantities. Since then, many theoretical proposals to describe gravity with a variable coupling has emerged, with scalar-tensor theories being the most famous ones (see, for instance, Ref.~\cite{Uzan:2010pm}). In the 1970s the possibility of a varying $G$ was discussed in the geophysical context to explain a possible expansion rate of the Earth \cite{1973QJRAS..14....9W}. Notwithstanding, most theories violating the strong equivalence principle will probably result in a time variation of its coupling constant \cite{Will:1993ns}. It is clear that any model with a varying $G$ must be consistent with experiments, with current strongest constraint being about $\dot{G}/G\lesssim 10^{-13}\,\mbox{yr}^{-1}$ \cite{PhysRevLett.93.261101}. The search for tiny variations of physical constants is a very active field and some attempts include Refs. \cite{Pasteka:2015hla,Pinho:2016mkm,Oreshkina:2017rue,Negrelli:2018toq,Safronova:2019lex,Martins:2019ebg,Giani:2020fpz}   

Essentially, all proposals of a varying $G$ coupling are formulated within the relativistic/covariant context. There is no consistent Newtonian theory admitting a varying $G$. The reason for that lies essentially on the non-trivial definition of energy conservation and issues related to invariance under Galilean transformations. Attempts to construct a Newtonian cosmology based on time-varying $G$ include contributions by Landsberg \& Bishop \cite{Landsberg:1975}, McVittie \cite{McVittie:1978} and Duval, Gibbons \& Horvathy \cite{Duval:1990hj} (see also Ref. \cite{10.1093/mnrasl/sly118}). Solutions of the McVittie proposal have been explored in Ref.\cite{Barrow:1996xn}. The main idea behind these approaches is to replace $G$ by $G(t)$ at the level of theory's dynamical equations. Consequently, one still needs some ansatz for $G(t)$, usually assuming a power-law dependence $G\propto t^{n}$. Thus, the evolution of the gravitational coupling is not obtained from a more fundamental aspect of the theory, but imposed by hand, instead. To the best of our knowledge there is no varying $G$ Newtonian gravity formulated from the classical Lagrangian formalism. This is the gap this work aims to fulfil.

Brans \& Dicke elaborated a very elegant prototype of a covariant scalar-tensor gravitational theory in which the gravitational coupling is a regular function of the scalar field $\phi$ \cite{Brans:1961sx}. Therefore, the dynamical evolution of the scalar field naturally induces the running of the gravitational coupling. Inspired by the Brans-Dicke covariant theory, here we propose a gravitational Lagrangian yielding to a varying $G$ theory. 

Of course, one can ask about the meaning of a Newtonian-type theory with a varying $G$. The answer can be given in many different ways. First, the problem itself is interesting since it can shed light on the specificity and the structure of the Newtonian theory of gravity, the oldest scientific theory to describe the gravitational phenomena, which has been very successful until the emergence of Einstein's general relativity. We must recall that many phenomena in small scales (scales of the solar system, stars, galaxy, cluster of galaxies, etc) can be treated using the Newtonian framework.
In this sense, if $G$ is not constant it can be useful to have a Newtonian theory incorporating this feature. It is also important to remember that most of the tests on the constancy of $G$ are done using systems for which the Newtonian approximation is valid. 
Moreover, many numerical simulations in cosmology employ the Newtonian theoretical framework too, and a modified formalism including effects of a non-constant $G$ could be equally fruitful in this area.

We start by reviewing the standard gravitational Lagrangian within the Newtonian theory in Section \ref{sec:newtonian}. It is also discussed the consequences for a static and spherically symmetric source with a constant matter density, as well as the cosmological background cases. These shall be useful for further comparison with results obtained in Section \ref{sec:classical_g} where we present our Brans-Dicke inspired varying $G$ Newtonian gravity theory. To finish, we draw our conclusions in Section \ref{sec:conclusion}.

\section{The standard Newtonian gravity} \label{sec:newtonian}

The Poisson equation of the Newtonian theory of gravity can be obtained from the Lagrangian,  
\begin{eqnarray}
\label{poisson}
{\cal L} = \frac{\nabla\psi_{\rm\s N}\cdot\nabla\psi_{\rm\s N}}{2} + 4\pi G \rho \psi_{\rm\s N},
\end{eqnarray}
where $\rho$ is the matter density, $\psi_{\rm\s N}$ is the Newtonian gravitational potential and $G$ is the gravitational coupling.
Inserting this Lagrangian in the Euler-Lagrange equation,
\begin{eqnarray}
\nabla\cdot\frac{\partial{\cal L}}{\partial \nabla\psi_{\rm\s N}} - \frac{\partial{\cal L}}{\partial\psi_{\rm\s N}} = 0.
\end{eqnarray} 
the Poisson equation is directly obtained,
\begin{eqnarray}
\nabla^2\psi_{\rm\s N} = 4\pi G\rho.
\end{eqnarray}
For a point mass, Poisson's equation implies that the gravitational force depends on the inverse square of the distance between the position of the source and a test particle.

\subsection{Static and spherically symmetric solution for a homogeneous mass distribution}

Let us review a very simple application of the Poisson equation: the computation of the gravitational potential of a homogeneous sphere of radius $R$ and mass $M$, with a constant density $\rho_0$, i.e.,
\begin{align}
\rho &= \rho_0,\quad  \mbox{for} \  0 < r \leq R,\\
\rho &= 0,\quad \mbox{for} \ r > R.
\end{align}
For the exterior region $r > 0$, the solution reads
\begin{eqnarray}
\psi_{\rm\s N}(r) = A + \frac{B}{r},
\end{eqnarray}
where $A$ and $B$ are arbitrary integral constants. Setting the gravitational potential to vanish at infinity, as usual, will lead to $A=0$. In the interior region $0 < r \leq R$,  Eq. (\ref{poisson}) reduces to,
\begin{eqnarray}
\psi_{\rm\s N}'' + 2 \frac{\psi_{\rm\s N}'}{r} = 4\pi G\rho,
\end{eqnarray}
with the upper prime indicating a derivative with respect to $r$. The solution of the above equation is given by,
\begin{eqnarray}
\psi_{\rm\s N}(r) = \frac{4\pi G\rho_0 r^2}{6} + C + \frac{D}{r},
\end{eqnarray}
with $C$ and $D$ being integral constants.
The regularity of the potential at the origin demands that $D = 0$. Moreover, junction conditions at $r = R$ leads to $B=-GM$ and $C=-3GM/2R$, after using $\rho_0 = 3M/4\pi R^3$, where $M$ is the total mass of the source. The final solution reads,
\begin{align}
\psi_{\rm\s N}(r) =& \frac{GMr^2}{2R^3} - \frac{3GM}{2R}, \qquad \mbox{for}\quad 0 \leq r \leq R, \\[1ex]
\psi_{\rm\s N}(r) =& - \frac{GM}{r}, \qquad \mbox{for}\quad r > R.
\end{align}

\subsection{Cosmology in the standard Newtonian theory}

In order to study a cosmological scenario using Newtonian theory the most direct approach is to consider the universe
as a homogeneous and isotropic expanding self-gravitating fluid \cite{Milne:1934, McCrea:1934}. Hence, the fundamental set of equations is formed by the continuity equation (expressing the conservation of matter), the Euler equation (Newton's second law expressed in a convenient way to study fluids) and the Poisson equation,
\begin{eqnarray}
\frac{\partial\rho}{\partial t} + \nabla\cdot(\rho \vec v) = 0,\\
\frac{\partial\vec v}{\partial t} + \vec v\cdot\nabla\vec v = - \frac{\nabla p}{\rho} - \nabla\psi_{\rm\s N},\\
\nabla^2\psi_{\rm\s N} = 4\pi G\rho.
\end{eqnarray}
The density $\rho$ and the pressure $p$ depend only on time. In order to take into account the cosmological expanding background via the Hubble-Lemaître law, the velocity field is written as,
\begin{eqnarray}
\vec v = \frac{\dot{a}}{a}\,\vec r,
\end{eqnarray}
where $a$ is a given function of time, which in the relativistic context represents the scale factor.

For the pressureless case, mimicking a cosmological matter dominated epoch, the above equations have the following solutions,
\begin{eqnarray}
\rho = \rho_0\,a^{-3},\\[1ex]
a(t) = a_0\,t^{2/3},
\end{eqnarray}
with $a_0$ being a constant.
These solutions are equivalent to the ones obtained with Einstein's general relativity for in the case of a universe filled with pressureless matter \cite{Weinberg:1972kfs}.

\section{Newtonian theory with variable \textit{G}}\label{sec:classical_g}

Now, we want to design a classical theory with varying gravitational coupling. Of course, the proposed theory intends to result in a consistent scenario for typical self-gravitating systems e.g., cosmology, stars, etc. Inspired by the Brans-Dicke recipe to construct a relativistic gravitational theory with a varying gravitational coupling, we propose the following Lagrangian:
\begin{equation}\label{LagrangianG}
{\cal L} = \frac{\nabla\psi\cdot\nabla\psi}{2} - \frac{\omega}{2}\biggr(\psi\frac{\dot\sigma^2}{\sigma^2} - c^4\nabla\sigma\cdot\nabla\sigma \biggl) + 4\pi G_0 \rho \sigma \psi,
\end{equation}
where $\sigma$ is a new function related to the gravitational coupling. It can depend on both space and time. Also, we have introduced the parameter $\omega$ which shall be assumed constant henceforth. Once we are now about to have a varying-$G$ theory, we define $G_0$ to represent a truly constant term, which is related with Newton's gravitational constant as will become clear later on.

In some sense the Lagrangian above corresponds to the Newtonian version of the relativistic Brans-Dicke theory (in Einstein's frame). The constant $c$ appears in this Lagrangian for dimensional reasons. This does not mean this is a relativistic theory since this Lagrangian is invariant under the Galilean group transformations. At this level, the introduction of the velocity of light, dictated by dimensional reasons, may be viewed as a consequence of another classical theory, the electromagnetism, with its two fundamental constants, the electric permittivity $\epsilon_0$ and magnetic permeability $\mu_0$ in vacuum which is understood as the absence of usual atomic matter. If we do not want to evoke a relativistic principle, this quantity, derived from $\epsilon_0$ and $\mu_0$ is the only possible fundamental constant with dimension of velocity. Of course, if want to mention the limit velocity established by Special Relativity (SR) such remark may look completely idle. Such explicit mention to SR is what we want to avoid in order to keep our reasoning, as far as possible, in a non-relativistic context.

Applying the Euler-Lagrangian equations of motion, 
\begin{align}
\nabla\cdot\frac{\partial{\cal L}}{\partial \nabla\psi} - \frac{\partial{\cal L}}{\partial\psi} &= 0,\\
\frac{d}{dt}\frac{\partial{\cal L}}{\partial\dot\sigma} + \nabla\cdot\frac{\partial{\cal L}}{\partial \nabla\sigma} - \frac{\partial{\cal L}}{\partial\sigma} &= 0,
\end{align}
the following equations are obtained:
\begin{align}
\label{en1}
\nabla^2\psi + \frac{\omega}{2}\left(\frac{\dot\sigma}{\sigma}\right)^2  &= 4\pi G_0 \sigma \rho,\\
\label{en2}
c^4\frac{\sigma}{\psi}\nabla^2\sigma - \frac{d}{dt}\biggr(\frac{\dot\sigma}{\sigma}\biggl) - \frac{\dot\psi}{\psi}\frac{\dot\sigma}{\sigma} &= \frac{4\pi G_0\sigma \rho}{\omega}.
\end{align}
The over-dot indicate total time derivative, which assures to the resulting equations an invariance with respect to Galilean transformations. Equations \eqref{en1}-\eqref{en2} show explicitly that the effective gravitational constant is given by $G_0\sigma$.
As we will verify later, the standard Newtonian limit is recovered when $\sigma$ is constant and $\omega \rightarrow \infty$. These same conditions lead Brans-Dicke theory to general relativity.

\subsection{Static and spherically symmetric solution for a homogeneous mass distribution}

Even if the set of equations (\ref{en1}) and (\ref{en2}) can not be trivially solved one might expect that the they lead to a modification of the usual Newtonian gravitational force. We
can have a more clear picture of such deviation by studying the static spherically symmetric mass distribution with
constant density as it has been carried out in the previous section.

By considering a static sphere of radius $R$ with constant density $\rho_0$ and assuming henceforth that $\omega>0$ (we comment on the $\omega<0$ case at the end of this section) the equations (\ref{en1}) and (\ref{en2}) reduce to,
\begin{equation}
\label{en3}
\nabla^2\psi  = \frac{4\pi G_0\rho_0}{c^2\sqrt{\omega}}\,\tilde\sigma ,
\end{equation}
\begin{equation}
\label{en4}
\nabla^2\tilde\sigma = \frac{4\pi G_0\rho_0}{c^2\sqrt{\omega}}\,\psi,
\end{equation}
with $\tilde\sigma = c^2\sqrt{\omega}\,\sigma$. These equations can be combined such that,
\begin{eqnarray}\label{eq-phi-sigma}
\psi\nabla^2\psi - \tilde\sigma\nabla^2\tilde\sigma = 0.
\end{eqnarray}
One possible solution is,
\begin{eqnarray}\label{sol}
\tilde\sigma = \pm \psi \quad \Rightarrow \quad \sigma = \pm\frac{\psi}{c^2\sqrt{\omega}}.
\end{eqnarray}
We will chose the upper sign in relations (\ref{sol}).
Then, Eq. (\ref{en3}) becomes a homogeneous modified Helmholtz equation,
\begin{eqnarray}\label{helmholtz}
\nabla^2\psi - k^2\psi=0,
\end{eqnarray}
where we have defined,
\begin{eqnarray}
k^2 = \frac{4\pi G_0\rho_0}{c^2\sqrt{\omega}}.
\end{eqnarray}
Note that, in the exterior region $r > R$, where $k=0$, we continue to have a Laplace's equation. Then, the solution is the same as in the Newtonian standard case,
\begin{eqnarray}
\psi = A + \frac{B}{r}.
\end{eqnarray}
For the interior mass distribution $r < R$, the general solution of \eqref{helmholtz} is given by, 
\begin{eqnarray}
\psi = \frac{1}{\sqrt{r}}\biggr\{CK_{1/2}(kr) + DI_{1/2}(kr)\biggl\},
\end{eqnarray}
where $K_\nu(z)$ and $I_\nu(z)$ are the modified Bessel functions of first and second rank, respectively. The function $K_\nu(z)$ is singular at the origin and it must be discarded, thus we set $C=0$.
Moreover, remark that
\begin{eqnarray}
I_{1/2}(z) = \sqrt{\frac{2z}{\pi}}\frac{\sinh{z}}{z}.
\end{eqnarray}
Imposing the matching conditions at $r = R$, it comes out,
\begin{align}
D &= \frac{-B\sqrt{k\pi/2}}{kR\cosh(kR) - \sinh(kR)},\\
A &= \frac{-kB}{kR - \tanh(kR)}.
\end{align}

Note that, with $A\neq 0$, one would have a gravitational potential that does not vanish at infinity. However, the dynamical equation \eqref{eq-phi-sigma}, together with the ansatz \eqref{sol}, is invariant under the transformations $\psi\rightarrow\psi + \lambda$ and $\tilde\sigma\rightarrow\tilde\sigma+\lambda$, with $\lambda$ a constant. One can then work with $\lambda=-A$ to obtain the following configuration for the gravitational potential,
\begin{widetext}
\begin{align}
    \psi(r) &= -\frac{G_0M\,k}{kR\cosh{(kR)}-\sinh{(kR)}}\left[\cosh(kR)-\frac{\sinh{(k\,r)}}{k\,r}\right],\qquad \mbox{for}\quad r<R,\label{psi-int}\\[1ex]
    \psi(r) &= -\frac{G_0M}{r},\qquad \mbox{for}\quad r\geq R.
\end{align}
\end{widetext}
In the above expressions we have already make the identification $B = - G_0M$, such that the exterior gravitational force, acting on a test particle, matches the Newtonian one. One then can see that $G_0$ must match the value of Newton's gravitational constant.
This is a general property for any static vacuum configuration, since \eqref{en1} always assume a Laplace's equation when the $\psi$ and $\sigma$ are time independent and $\rho=0$. Thus, this varying-$G$ Newtonian theory introduces effective modifications only inside matter. It is worth to note that Palatini formulations of $f(R)$ theories also introduce modifications of Newtonian gravity only inside matter content.

\begin{figure*}[!t]
\begin{subfigure}{.48\textwidth}
\centering
\includegraphics[width=\linewidth]{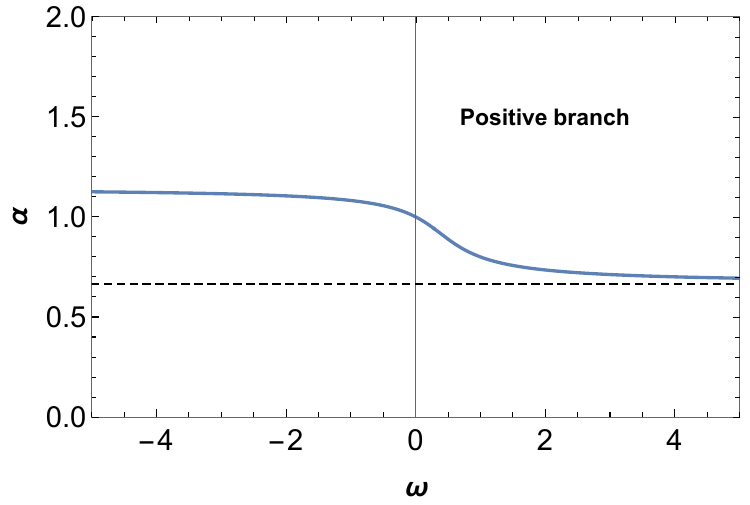}
\end{subfigure}
\begin{subfigure}{.48\textwidth}
\centering
\includegraphics[width=\linewidth]{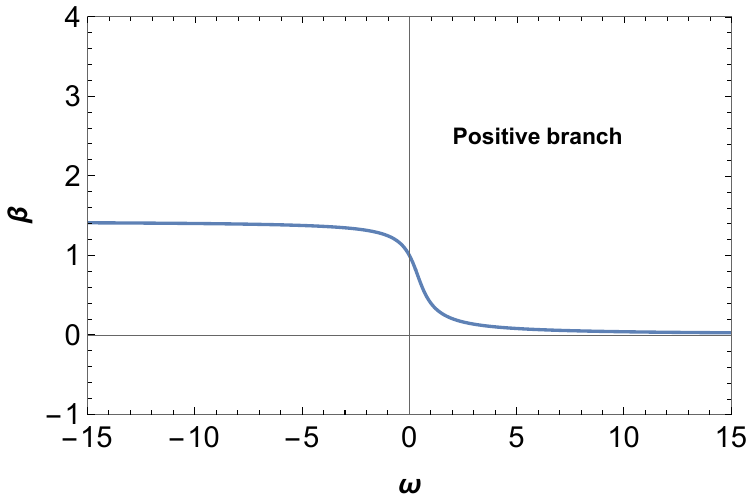}
\end{subfigure}
\caption{Dependence of the cosmological parameters $\alpha$ and $\beta$ with $\omega$ taking the positive branch in solution (\ref{ae}). Horizontal dashed line in left panel denotes the value $\alpha=2/3$.}
\label{fig:1}
\end{figure*}

However, the Yukawa-like potential for the interior solution is a particular behavior for the constant matter density configuration. When $\rho$ is a function of spatial coordinates, Eq.~\eqref{helmholtz} will not be a Helmholtz equation anymore. One thus need to investigate more realistic star configurations in order to better understand the modifications brought by this model. We leave this for a future work. Even so, for this simplistic model, one would not expect great deviations from the Newtonian interior potential. This because $k$ would be a typical small quantity. In fact, one can write,
\begin{equation}
    kr\sim \frac{10^{-2}}{\omega^{1/4}}\sqrt{\frac{M/M_\odot}{R/R_\odot}}\,\left(\frac{r}{R}\right),
\end{equation}
where $M_\odot$ and $R_\odot$ are the mass and radius of the Sun. It is then possible to estimate that significant deviations would only appear for (constant density) star configurations having,
\begin{equation}\label{MtoR}
    \frac{M/M_\odot}{R/R_\odot}\gtrsim 10^4\sqrt{\omega}.
\end{equation}
Thus, for very compact objects and small values of $\omega$ there could appear departures from the standard Newtonian physics inside matter regions. Assuming that $kR\ll 1$, one can verify that expression \eqref{psi-int} tends to the Newtonian potential,
\begin{equation}
    \psi(r\!<\!R)\approx \psi_{\rm\s N}(r\!<\!R) + O(k^2).
\end{equation}
One can expect then that a small discrepancy occurs near the origin for small $\omega$ values.

For $\omega<0$ relation (\ref{sol}) is not valid anymore. However, Eqs. (\ref{en3}) and (\ref{en4}) can be combined leading to a fourth order differential equation which admits solutions in terms of a combination of trigonometric and hyperbolic functions. By choosing a regular solution near the origin it is possible to show that similar restrictions to (\ref{MtoR}) are found. Therefore, both cases are essentially the same.

\subsection{Cosmology in the varying \textit{G} Newtonian gravity}

Let us turn now our attention to the cosmological case. Since a spatial dependence of $\sigma$ would be inconsistent with a pure time dependence of the density $\rho$, see Eqs. (\ref{en1})-(\ref{en2}), we also assume a temporal dependence for $\sigma$. A pure time dependent $\sigma$ is also in agreement with the original Dirac's proposal. On the other hand, as in the cosmological set using standard Newtonian theory, the gravitational potential is considered to be a function of both time and spatial coordinates.
Under these conditions the dynamical equations become,
\begin{align}
\label{enc1}
\nabla^2\psi + \frac{\omega}{2}\frac{\dot\sigma^2}{\sigma^2}= 4\pi G_0 \sigma \rho ,\\
\label{enc2}
\frac{d}{dt}\biggr(\frac{\dot\sigma}{\sigma}\biggl) + \frac{\dot\psi}{\psi}\frac{\dot\sigma}{\sigma} = - \frac{4\pi G_0\sigma \rho}{\omega}.
\end{align}

While the Poisson equation (\ref{enc1}) remains the same compared with the static case, the equation for the dynamical evolution of $\sigma$ is different.  
From (\ref{enc1}), one can write
\begin{eqnarray}\label{43}
\psi = \frac{4\pi G_0\sigma\rho}{6}r^2 -  \frac{\omega}{12}\frac{\dot\sigma^2}{\sigma^2} r^2.
\end{eqnarray}

It is appropriate to parametrize the cosmological evolution in terms of the scale factor $a(t)$. Let us look for power law solutions under the form,
\begin{eqnarray}
a = a_0t^{\alpha},  \quad \sigma = \sigma_0t^{\beta},
\end{eqnarray}
with $\alpha$, $\beta$, $a_0$ and $\sigma_0$ constants. From the conservation law we have, as in the Newtonian standard case,
\begin{eqnarray}
\rho = \rho_0a^{-3}.
\end{eqnarray}
By considering power law solutions for $\psi$ and taking into account Eq. (\ref{43}), the potential $\psi$ must take the form,
\begin{eqnarray}\label{psicosmo}
\psi = \frac{\psi_0}{t^2}\,r^2,
\end{eqnarray}
where,
\begin{eqnarray}
\psi_0 = \frac{4\pi G_0\rho_0\sigma_0}{6a_0^3} - \frac{\omega}{12}{\beta}^2.
\end{eqnarray}
For the same reason, the coefficients $\alpha$ and $\beta$ must obey the relation,
\begin{eqnarray}
\beta = - 2 + 3 \alpha.
\end{eqnarray}
As expected, it is worth noting that for $\alpha = 2/3$ (mimicking a cosmological dust matter expansion phase) one finds $\beta = 0$, i.e., the gravitational coupling becomes constant.     



\begin{figure*}[t]
\begin{subfigure}{.48\textwidth}
\centering
\includegraphics[width=\linewidth]{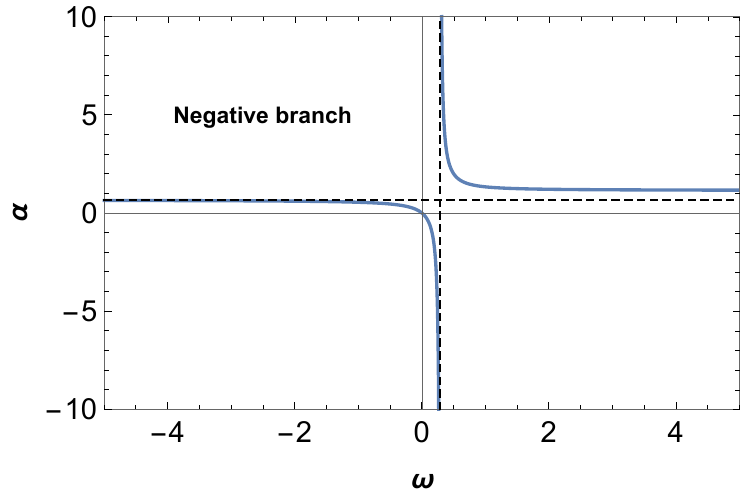}
\end{subfigure}
\begin{subfigure}{.48\textwidth}
\centering
\includegraphics[width=\linewidth]{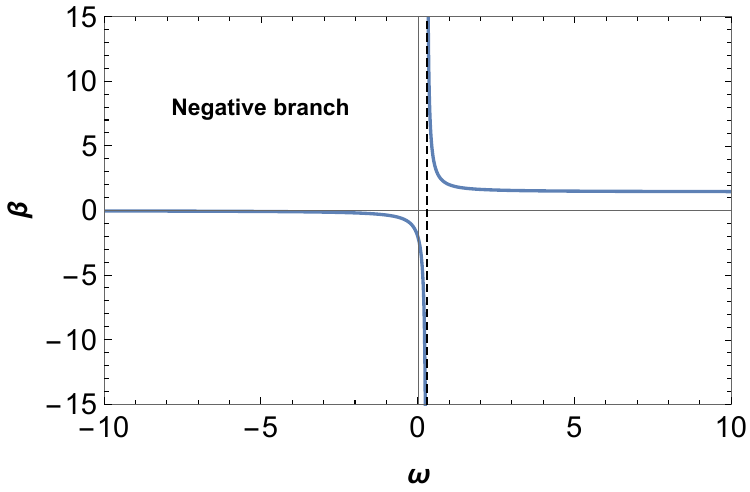}
\end{subfigure}
\caption{Dependence of the cosmological parameters $\alpha$ and $\beta$ with $\omega$ taking the negative branch in solution (\ref{ae}). Horizontal dashed line in the left panel denotes the value $\alpha=2/3$. In both panels the vertical dashed line denotes $\omega=2/7$.}
\label{fig:2}
\end{figure*}

Let us take into account the Euler equation with $\psi$ giving by (\ref{psicosmo}) and expressing the velocity field according to the Hubble-Lemaître law we obtain,
\begin{eqnarray}\label{modfriedmann}
\frac{\ddot a}{a} = - \frac{4\pi G_0\sigma\rho}{3} + \frac{\omega}{6}\biggr(\frac{\dot\sigma}{\sigma}\biggl)^2.
\end{eqnarray}
Using the relations previously found and Eq. \eqref{enc2}, we find that the parameter $\alpha$ must obey a second order algebraic equation:
\begin{eqnarray}
\biggr(6 - 21\omega\bigg)\alpha^2 - (6 - 38\omega)\alpha  - 16\omega = 0,\label{alpha-eq}
\end{eqnarray}
with solution,
\begin{eqnarray}
\label{ae}
\alpha = \frac{3 - 19\omega \pm \sqrt{25\omega^2 -18\omega + 9}}{6 - 21\omega}.
\end{eqnarray}

In Figs. (\ref{fig:1}) and (\ref{fig:2}) the dependence of parameters $\alpha$ and $\beta$ with $\omega$ are displayed for the two possible values of $\alpha$ according to the sign in (\ref{ae}). It is worth noting that the usual Newtonian limit is obtained, for the upper sign in (\ref{ae}) when $\omega \rightarrow \infty$, while for the lower sign it happens at $\omega \rightarrow - \infty$. As in Brans-Dicke theory, Newtonian theory is recovered only when $\omega$ is very large and also $\sigma$ is constant. For the upper sign, Fig. (\ref{fig:1}), the gravitational coupling always increase $(\beta>0)$, and $\alpha \geq 2/3$, with the universe having an accelerated expansion ($\alpha > 1$) when $\omega < 0$. In the lower sign branch, Fig. (\ref{fig:2}), the situation is more involved: for $\omega > 2/7$, $\alpha > 1$ and $\beta > 0$, i.e. the gravitational coupling is increasing while the universe expands accelerated; but if $\omega < 2/7$, the gravitational coupling is always decreasing implying a contracting universe for $0 < \omega < 2/ 7$, or a decelerated expanding universe for $\omega < 0$. The critical value $\omega = 2/7$ implies that Eq. \eqref{alpha-eq} is linear, with single solution $\alpha=16/17$.

\section{Conclusions}\label{sec:conclusion}
A considerable number of varying $G$ gravitational theories have been proposed along the last century. Most of them rely upon a covariant description, spanning from the first prototype of scalar-tensor theories, idealized by Brans and Dicke \cite{Brans:1961sx}, to the modern Horndeski theories \cite{Horndeski:1974wa}. At the Newtonian level all attempts so far relied on the {\it ad hoc} introduction of a varying $G$ as e.g., in the McVittie proposal \cite{McVittie:1978}. This work introduces a Newtonian formulation for a gravitational theory in which the variation (both temporal and spatial) of $G$ emerges naturally from the Lagrangian formalism (\ref{LagrangianG}). 

We have found that the gravitational potential of the proposed theory is always equivalent to the Newtonian one in vacuum. Inside matter distributions deviations are negligible for ordinary mass-radius rates, unless the  parameter $\omega$ assumes very small values. Our analysis has been performed for the simplified case of a constant density spherically symmetric object and more involved configurations should be studied in the near future. Moreover, the entire formalism recovers the standard Newtonian results in the limit $|\omega| \rightarrow \infty$ and $\sigma \rightarrow$ constant similarly to the covariant Brans-Dicke theory. However it is worth noting that though our proposal has its structure inspired by the Brans-Dicke theory one can not obtain it, in the appropriate limit, from the latter. Strictly speaking, our approach can not be seen as a Newtonian Brans-Dicke theory since locally the weak-field limit of the original Brans-Dicke theory implies in a constant gravitational coupling.

 
The cosmological framework shows a clear contribution of the variation of the field $\sigma$ as seen in the modified Friedmann equation (\ref{modfriedmann}). The background expansion depends on the $\omega$ value converging to the Einstein-de Sitter expansion (as in a pure dust general relativity model) in the limit $\omega \rightarrow \infty$.
In special, when the gravitational coupling grows with time the expansion rate is enhanced; in particular cosmic accelerated expansion is allowed due to the growing of the gravitational coupling. Then, although not explored here, a dynamical evolution of $\omega$ could explain the transition to the accelerated cosmological expansion phase associated to dark energy.

The possibility of a dynamical $\omega$ parameter and its value should be investigated in future works by studying the stellar interior via a modified Lane-Emden equations as well as using cosmological data.

\begin{acknowledgments}
We dedicate this work to the memory of Antonio Brasil Batista, one the founders of the research group Cosmo-ufes, who has introduced us to the problem of the variation of the gravitational coupling in newtonian and relativistic theories. We thank Davi C. Rodrigues for enlightening discussions on the subject of this paper. JCF thanks CNPq and FAPES for partial financial support.
HV thanks CNPq and PROPP/UFOP for partial financial support. JDT thanks FAPES and CAPES for their support through the Profix program. TG thanks FAPES for their support. 
\end{acknowledgments}

\bibliography{Refs}

\end{document}